\documentclass[aps,nofootinbib,superscriptaddress,twocolumn]{revtex4-1}
\usepackage{graphicx}
\usepackage{amsmath,amssymb}
\usepackage{hyperref}
\usepackage[usenames]{color}

\font \bolditalics = cmmib10
\def\bx#1{\leavevmode\thinspace\hbox{\vrule\vtop{\vbox{\hrule\kern1pt
        \hbox{\vphantom{\tt/}\thinspace{\bf#1}\thinspace}}
      \kern1pt\hrule}\vrule}\thinspace}
\def \vc #1{{\textfont1=\bolditalics \hbox{$\bf#1$}}}

\def\thetavc{{\vc \theta}}

\hypersetup{
    colorlinks=true,
    linkcolor=red,
    citecolor=blue,
}

\def\be{\begin{equation}}
\def\ee{\end{equation}}
\def\ba{\begin{eqnarray}}
\def\ea{\end{eqnarray}}

\def \msun{{\rm M}_{\odot}}

\begin{document}

\title{Dissecting the thermal Sunyaev-Zeldovich-gravitational lensing cross-correlation with hydrodynamical simulations}

\author{Alireza Hojjati}
\affiliation{Department of Physics and Astronomy, University of British Columbia, Vancouver, V6T 1Z1, BC, Canada}
\affiliation{Physics Department, Simon Fraser University, Burnaby, V5A 1S5, BC, Canada}

\author{Ian G. McCarthy}
\affiliation{Astrophysics Research Institute, Liverpool John Moores University, Liverpool, L3 5RF, United Kingdom}

\author{Joachim Harnois-Deraps}
\affiliation{Department of Physics and Astronomy, University of British Columbia, Vancouver, V6T 1Z1, BC, Canada}
\affiliation{Canadian Institute for Theoretical Astrophysics, University of Toronto, M5S 3H8, ON, Canada}

\author{Yin-Zhe Ma}
\affiliation{Jodrell Bank Centre for Astrophysics, School of Physics and Astronomy, University of Manchester, Oxford Road, Manchester M13 9PL, United Kingdom}

\author{Ludovic Van Waerbeke}
\affiliation{Department of Physics and Astronomy, University of British Columbia, Vancouver, V6T 1Z1, BC, Canada}
\affiliation{Canadian Institute for Advanced Research, 180 Dundas St W, Toronto, ON M5G 1Z8, Canada}

\author{Gary Hinshaw}
\affiliation{Department of Physics and Astronomy, University of British Columbia, Vancouver, V6T 1Z1, BC, Canada}
\affiliation{Canadian Institute for Advanced Research, 180 Dundas St W, Toronto, ON M5G 1Z8, Canada}
\affiliation{Canada Research Chair in Observational Cosmology}

\author{Amandine M. C. Le Brun}
\affiliation{Astrophysics Research Institute, Liverpool John Moores University, Liverpool, L3 5RF, United Kingdom}
\affiliation{DSM/Irfu/SAp, CEA Saclay, 91191 Gif-sur-Yvette, France}

\begin{abstract}
We use the cosmo-OWLS suite of cosmological hydrodynamical simulations, which includes different galactic feedback models, to predict the cross-correlation signal between weak gravitational lensing and the thermal Sunyaev-Zeldovich (tSZ) $y$-parameter. The predictions are compared to the recent detection reported by van Waerbeke and collaborators.  The simulations reproduce the weak lensing-tSZ cross-correlation, $\xi_{y\kappa}(\theta)$, well.  The uncertainty arising from different possible feedback models appears to be important on small scales only ($\theta \lesssim 10$ arcmin),  while the amplitude of the correlation on all scales is sensitive to cosmological parameters that control the growth rate of structure (such as $\sigma_8$, $\Omega_m$ and $\Omega_b$).  This study confirms our previous claim (in Ma et al.) that a significant proportion of the signal originates from the diffuse gas component in low-mass ($M_{\rm{halo}} \lesssim 10^{14} \msun$) clusters as well as from the region beyond the virial radius.  We estimate that approximately 20$\%$ of the detected signal comes from low-mass clusters, which corresponds to about 30$\%$ of the baryon density of the Universe.  The simulations also suggest that more than half of the baryons in the Universe are in the form of diffuse gas outside halos ($\gtrsim 5$ times the virial radius) which is not hot or dense enough to produce a significant tSZ signal or be observed by X-ray experiments.  Finally, we show that future high-resolution tSZ-lensing cross-correlation observations will serve as a powerful tool for discriminating between different galactic feedback models.
\end{abstract}

\maketitle

\section{Introduction}
To obtain a complete understanding of structure formation in the Universe we need to better understand the evolution of baryons on large scales.  Only about 10\% of all the baryons in the Universe reside in stars and cold gas in galaxies \cite{Balogh2001,Budzynski2014} while the rest is thought to reside in a diffuse gas component spread over a wide range of scales, densities and temperatures. Observing this component is difficult and is presently limited to regions where the gas is hot and dense, where it can be be detected via X-ray emission and/or the thermal Sunyaev-Zeldovich (tSZ) effect.  These conditions are typically limited to the central parts of massive halos; thus we have few constraints on diffuse gas beyond the virial radius and/or in low mass halos ($M_{\rm halo} \lesssim 10^{14} \msun$).  

The apparent deficit of baryons in massive halos \cite{Bregman2007,Sun2009}, combined with the observation that energetic winds are ubiquitous in high-$z$ galaxies \cite{Nesvadba2008} suggests that feedback must be efficient at ejecting baryons from their halos.  Thus, important and independent constraints on models of feedback can be obtained by observing the diffuse gas outside halos.  Indeed, getting the feedback model(s) right is absolutely crucial for our understanding of galaxy formation and for our ability to use galaxies and haloes to constrain cosmology.
 

Below $z\sim6$ the majority of hydrogen in the Universe is ionized.  Observations of atomic emission lines can probe the warm ionized gas \cite{Shull:2011aa} as well as the already-noted X-ray and tSZ probes.  The latter observations are well suited to probe low density environments but current observations are limited by sensitivity and/or angular resolution, and by confusion from Galactic and extragalactic dust emission. However, the {\it cross-correlation} of $y$ with gravitational lensing allows us to selectively probe gas at lower $y$ than we could with the auto-correlation, and to directly compare the relationship between gas and mass. To date, only two such cross-correlations have been reported: \cite{CFHT1} measured a 6-$\sigma$ correlation between the tSZ effect and weak lensing convergence from relatively low-redshift lenses ($z\sim0.4$).  The authors used lensing maps derived from the Canada-France-Hawaii Telescope Lensing Survey (CFHTLenS, \cite{Heymansetal2012, Erbenetal2013}) and tSZ maps derived from the {\it Planck} satellite over a sky area of $\sim$150 deg$^2$. Assuming a constant bias between gas and total mass, $b_{\rm gas}$, they derived a joint constraint on the gas bias, density and temperature of: $(b_{\rm gas})(T_{\rm e} / 0.1 \textrm{ keV})(\bar{n}_{\rm e} / 1\textrm{ m}^{-3}) = 2.01 \pm 0.52$.  Using {\it Planck} data alone, \cite{Hill:2013dxa} reported a 6-$\sigma$ detection of the Cosmic Microwave Background (CMB) lensing-tSZ cross-correlation.  The latter study differs from the former in that the CMB lensing signal is mainly sensitive to lenses located at high redshift $z>2$.  Both studies concluded that hot, ionized gas approximately traces dark matter over a wide range of scales ($\sim 0.1$ to $50\,h^{-1}\,{\rm Mpc}$).

The interpretation of the cross-correlation measurements is still an open question: the data can probe gas in regions beyond $R_{500}$, and even beyond $R_{200}$ ($R_{\rm vir}$), where matter is not in hydrostatic equilibrium and accurate modeling of the gas could be an issue.  An interpretation of the signal measured in \cite{CFHT1} was carried out in \cite{CFHT2} using the halo model.  This study concluded that up to $\sim$40\% of the signal comes from low-mass halos, and that a significant fraction of  the baryons reside at halo radii beyond $R_{\rm vir}$.   Another finding was a possible tension between the cross-correlation signal and the `universal pressure profile' (hereafter UPP) \citep{Arnaud:2009tt} that might indicate that the UPP over-predicts the small-scale tSZ signal. Confirmation of this interpretation could have important implications for the study of galaxy formation and the role of galactic feedback.  Here we point out that the halo model of \cite{CFHT2} assumed a best-fit {\it Planck} cosmology \cite{PlanckXVI}.  As we show below, the amplitude of the tSZ-lensing cross-correlation is highly sensitive to variations in cosmological parameters that control the growth rate of clusters (particularly $\sigma_8$ and $\Omega_m$, the baryon fraction $\Omega_b$ is also relevant since it dictates how much gas is present).  Thus, an alternative interpretation of the tension reported in \cite{CFHT2} is a possible mild tension with the best-fit {\it Planck} cosmology.  We comment more on this possibility below.  

Many of the questions mentioned above can be addressed with cosmological hydrodynamical simulations.  It would be particularly interesting to challenge the cross-correlation measurements against realistic simulations and to investigate the contribution of baryons to the signal from haloes of different mass and size.
This paper is a follow up on \cite{CFHT1, CFHT2} from the perspective of simulations. Namely, we use several sets of hydro simulations with different baryonic feedback models to make a wide range of tSZ and convergence maps, and further investigate the findings of \cite{CFHT1, CFHT2}.  

The organization of the paper is as follows: in \S~\ref{sec:formalism}, we briefly review the theoretical background, describe the cosmological simulations and baryonic feedback models employed, and review the cross-correlation procedure and results from \cite{CFHT1}.  In \S~\ref{sec:results}, we present the cross-correlation results derived from the simulations and compare them to the measured signal.  We summarize our results in \S~\ref{sec:summary}.

\section{Formalism and Method}
\label{sec:formalism}

\subsection{Cross-correlation of weak lensing and tSZ}
\label{cross-correlation}

Following the notations in \cite{CFHT1}, the gravitational lensing convergence $\kappa(\thetavc)$ is given by
\be
\kappa(\thetavc) = \int_0^{w_{\rm H}} {\rm d}w \, W^\kappa(w) \, \delta_{\rm m}(\thetavc f_K(w),w),
\label{eq:kappamapdef}
\ee
where $\thetavc$ is the position angle on the sky,  $w(z)$ is the comoving radial distance to redshift $z$,  $w_{\rm H}$ is the distance to horizon, $W^\kappa(w)$ is the lensing kernel \cite{CFHT1},
\be
W^\kappa(w) = {3 \over 2} \Omega_{\rm m} \left({H_0 \over c}\right)^2 g(w) \, {f_K(w) \over a},
\ee
$\delta_{\rm m}(\thetavc f_K(w),w)$ is the 3-dimensional mass density contrast, $f_K(w)$ is the angular diameter distance at comoving distance $w$, 
and the function $g(w)$ depends on the source redshift distribution $p_{\rm S}(w)$ as
\be
g(w) = \int_w^{w_{\rm H}} {\rm d}w' \, p_{\rm S}(w') \, {f_K(w'-w) \over f_K(w')}.
\ee
The tSZ signal is due to inverse Compton scattering of CMB photons off hot electrons along the line-of-sight which results in a frequency-dependent variation in the CMB temperature,
\be
{\Delta T\over T_0} = y \, S_{\rm SZ}(x),
\ee
where $S_{\rm SZ}(x) = x\coth (x/2) - 4$ is the tSZ spectral dependence, given in terms of $x=h\nu/k_{\rm B} T_0$, $h$ is the Planck constant, $k_{\rm B}$ is the Boltzmann constant, and $T_0=2.725$ K is the CMB temperature \cite{Sunyaev:1970eu}. The quantity of interest in the calculations here is the Comptonization parameter, $y$,  given by the line-of-sight integral of the electron pressure:
\be
y(\thetavc) = \int_0^{w_{\rm H}} a \, {\rm d}w \, {k_{\rm B} \sigma_{\rm T} \over m_{\rm e} c^2} \, n_{\rm e} T_{\rm e},
\label{y}
\ee
where $\sigma_{\rm T}$ is the Thomson cross-section, and $n_{\rm e}(\thetavc f_K(w),w)$ and $T_{\rm e}(\thetavc f_K(w),w)$ are the 3-dimensional electron number density and temperature, respectively. 

For the analysis in this paper, we mainly work with the real space cross-correlation function,  $\xi_{y\kappa}(\theta)$:
\be
\xi_{y\kappa}(\theta) = \sum_{\ell} \left(\frac{2\ell + 1}{4\pi}\right ) C_{\ell}^{y\kappa} \, J_0(\theta) \, b^y_{\ell} \, b^{\kappa}_{\ell},
\label{eq:ykappaxi}
\ee
where $C_{\ell}^{y\kappa}$ is the $y-\kappa$ angular cross-power spectrum, 
\be
C_{\ell}^{y\kappa} = {1 \over 2 \ell +1} \sum_m y_{\ell m} \kappa_{\ell m}^\ast,
\label{ykappaCel}
\ee
and $b^y_{\ell}$ and $b^{\kappa}_{\ell}$ are the Gaussian smoothing transfer functions of the $\kappa$ and $y$ maps, respectively. 

%
%

\subsection{Simulations}

\begin{table*}[th]
\centering
\caption{Sub-grid physics of the baryon feedback models in the cosmo-OWLS runs.  Each model has been run in both the {\it WMAP}-7 and {\it Planck} cosmologies \cite{McCarthy2014}.}
\begin{tabular}{|l|l|l|l|l|l|l|}
         \hline
	Simulation & UV/X-ray background & Cooling & Star formation & SN feedback & AGN feedback & $\Delta T_{\rm heat}$ \\
	\hline
        NOCOOL & Yes & No & No & No & No & ...\\
        REF & Yes & Yes & Yes & Yes & No & ...\\
        AGN 8.0 & Yes & Yes & Yes & Yes & Yes & $10^{8.0}$ K\\
        AGN 8.5 & Yes & Yes & Yes & Yes & Yes & $10^{8.5}$ K\\
        AGN 8.7 & Yes & Yes & Yes & Yes & Yes & $10^{8.7}$ K\\
        \hline
\end{tabular}
\label{table:cosmo_owls}
\end{table*} 

For this study we employ the cosmo-OWLS suite of cosmological hydrodynamical simulations.  cosmo-OWLS is an extension of the OverWhelmingly Large Simulations project (OWLS; \cite{Schaye2010}) and has been designed with cluster cosmology and large scale-structure surveys in mind. A detailed description of the cosmo-OWLS simulations can be found in \cite{LeBrun2014,vanDaalen2013, McCarthy2014}.

The simulation suite was run with a significantly modified version of the Lagrangian TreePM-SPH code \textsc{gadget3} \cite{Springel:2005mi} developed for the OWLS project.  The suite consists of box-periodic hydrodynamical simulations, the largest of which have volumes of $(400 \ h^{-1} \ {\rm Mpc})^3$ and $1024^3$ baryon and dark matter particles.  The initial conditions are based on either the {\it WMAP}-7 or {\it Planck} cosmologies with \{$\Omega_{\rm m}$, $\Omega_{\rm b}$, $\Omega_{\Lambda}$, $\sigma_8$, $n_{\rm s}$, $h$\} = \{0.272, 0.0455, 0.728, 0.81, 0.967, 0.704\} and \{0.3175, 0.0490, 0.6825, 0.834, 0.9624, 0.6711\}, respectively. 

We use five different baryon models from the suite, as summarized in Table \ref{table:cosmo_owls} and described in detail in \cite{McCarthy2014} and in references therein. \textsc{NOCOOL} is a standard non-radiative (`adiabatic') model. \textsc{REF} is the OWLS reference model and  includes sub-grid prescriptions for star formation \cite{Schaye:2007ss}, metal-dependent radiative cooling \cite{Wiersma:2008cs}, stellar evolution, mass loss, chemical enrichment \cite{Wiersmaetal2009}, and a kinetic supernova feedback prescription \cite{Vecchia:2008kn}. The \textsc{AGN} models are built on the \textsc{REF} model and additionally include a prescription for black hole growth and feedback from active galactic nuclei \cite{BoothSchaye2009}. The three \textsc{AGN} models differ only in their choice of the key parameter of the AGN feedback model $\Delta T_{\rm heat}$, which is the temperature by which neighbouring gas is raised due to feedback.  Increasing the value of $\Delta T_{\rm heat}$ obviously results in more energetic feedback events, but it also leads to more bursty feedback, since the black holes must accrete more matter in order to heat neighbouring gas to a higher adiabat.

We note here that none of the different baryon models have been calibrated to reproduce the observed properties of the hot gas in groups and clusters.  (Indeed, the main idea of cosmo-OWLS is to explore how the hot gas properties depend on uncertain subgrid physics.)  In spite  of this, \cite{LeBrun2014} have shown that the \textsc{AGN 8.0} model reproduces a wide range of X-ray and optical observations of local galaxy groups and clusters, and \cite{McCarthy2014} showed this model also reproduces the pressure distribution of the hot gas.  While the {\it abundance} of galaxy clusters depends strongly on the adopted cosmological parameter values, the internal structure and X-ray/optical scaling relations are only weakly dependent on cosmology (see \cite{LeBrun2014} for details). Neglect of AGN feedback (as in the \textsc{REF} model), on the other hand, leads to significant overcooling (i.e., excessive stellar mass fractions and star formation rates).  Increasing the heating temperature of the AGN feedback significantly higher than in the \textsc{AGN 8.0} model results in overly efficient ejection of gas from the progenitors of groups and clusters, resulting in hot gas mass fractions significantly lower than that inferred from X-ray-selected samples \cite{Sun2009}.   For these reasons we select \textsc{AGN 8.0} as the fiducial baryon feedback model and refer to it simply as \textsc{AGN}.  The other models are still useful, however, as they bracket the current observations.

Following \cite{McCarthy2014}, we produce light cones of the simulations by stacking randomly rotated and translated simulation snapshots along the line-of-sight back to $z=3$ \footnote{We use 15 snapshots at fixed redshift intervals between $z=0$ and $z=3$ for constructing the light cones.  This ensures a good comoving distance resolution needed to capture the evolution of the halo mass function and tSZ signal.}.  The light cones are used to produce $5^{\circ} \times 5^{\circ}$ $y$ and $\kappa$ maps. We construct 10 different light cone realizations for each feedback model and for the two background cosmologies.  Note that in the production of the $\kappa$ maps we adopt the source redshift distribution, $n(z)$, from the CFHTLenS survey (see \cite{Hildebrandtetal2012} and \cite{CFHT1} for details) to produce a consistent comparison with the observations.  

Although these are among the largest maps produced to date from light cones from self-consistent cosmological hydro simulations, they are still not sufficiently large to capture all of the ``ell-space'' modes relevant for the present study.  In Appendix~\ref{lowell}, we describe how we account for the finite viewing angle of the light cones when computing the simulated cross-correlation functions and test our methods via a comparison with the halo model.

The simulations we examine here were used in \cite{McCarthy2014} to predict the tSZ power spectra, $C_{\ell}^{yy}$, and the results were compared with the {\it Planck} measurements \cite{PlanckXXI}.  The authors found that spectra predicted with the {\it WMAP}-7 cosmology were in better agreement with the observations than those predicted using the {\it Planck} cosmological parameters.  Here we will investigate whether the same holds true for $C_{\ell}^{y\kappa}$.

\subsection{Gravitational lensing and tSZ data}

The details of the lensing and tSZ map making are given in \cite{CFHT1}; only the main results will be repeated here.  We use the gravitational lensing convergence maps from the CFHTLenS survey \cite{VanWaerbeke:2013eya}. The total area covered is $154$ deg$^2$ in four separate patches, and the maps are smoothed with a Gaussian window of $\theta_{\kappa, \rm FWHM}=10~{\rm arcmin}$ width. The mean lens redshift peaks at $z\sim 0.37$ (\cite{CFHT1}). 

Several full-sky maps of the Comptonization parameter, $y$, were constructed from the 15-month combined-survey {\it Planck} band maps.  Each map was constructed from a linear combination of four HFI frequency band maps (100, 143, 217, and 353 GHz)  and smoothed to a Gaussian beam profile with $\theta_{\rm SZ, FWHM} = 9.5$ arcmin.  The band coefficients were chosen such that the primary CMB signal is removed, and the dust emission with a spectral index $\beta_{\rm d}$ is nullified.  A range of $\beta_{\rm d}$ values were employed, resulting in a set of $y$ maps that were used as diagnostics of residual contamination.  The resulting $\xi_{y\kappa}$ measurements vary by roughly 10\% between the different $y$ maps, and we discuss this further below.

The cross-correlation measurements studied in this paper are identical to those reported in \cite{CFHT1}.  In the remainder of this paper we compare these measurements to the hydrodynamic simulations discussed above, first on large angular scale, then on small scales.

\section{Results}
\label{sec:results}

\subsection{Large-scale correlations}
\label{largescale}
 
\begin{figure}[th]
{\centering{
\includegraphics[width=1.0\columnwidth]{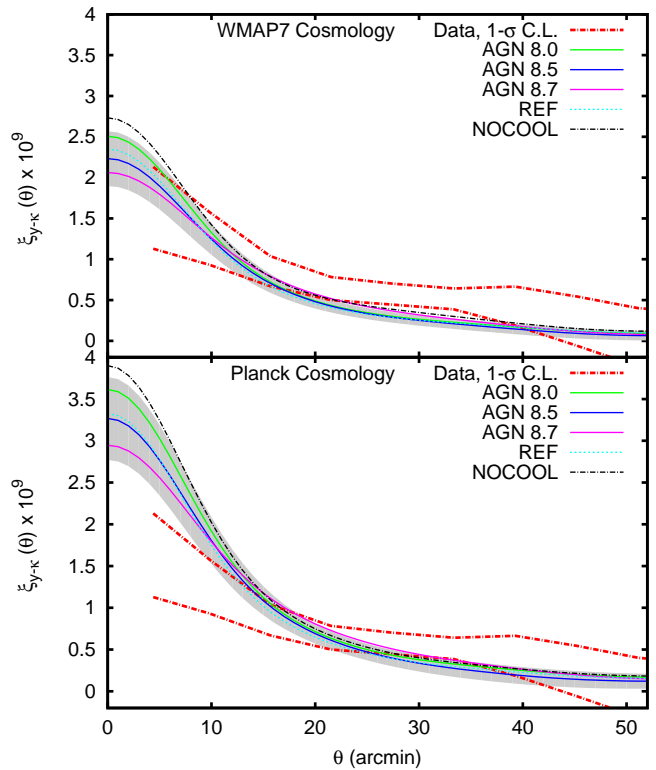}
}}
\caption{Comparison of the cross-correlation function $\xi_{y\kappa}$ computed from hydrodynamical simulations to the signal detected in \cite{CFHT1}.   Per Table~\ref{table:cosmo_owls}, five different baryon feedback models are considered.  The grey band represents the error on the mean value  of correlation function from each of the five feedback models in the simulations, rescaled to match the area covered by CFHTLenS, and averaged over the five models. The band is centred  over the \textsc{AGN 8.5} model. The red dashed-dotted lines represent the 1-$\sigma$ confidence interval on the measurement based on the different tSZ maps made in \cite{CFHT1} and includes statistical and systematic errors.  The simulations in the top panel use {\it WMAP}-7 cosmological parameters and the ones in the bottom panel use {\it Planck} parameters.}
\label{fig:Corr_y-kappa_smooth}
\end{figure}

Fig.~\ref{fig:Corr_y-kappa_smooth} compares simulated cross-correlation functions to the signal detected in \cite{CFHT1}.  Both panels show simulation results using different baryon feedback models; the top panel uses {\it WMAP}-7 cosmological parameters and the bottom panel uses {\it Planck} parameters.  The cross-correlation signal from the simulated maps is computed following Eq.~(\ref{eq:ykappaxi}), where the maps are smoothed to match the angular resolution of the {\it Planck} map and the CFHTLenS data. The grey band represents the error on the mean value derived from ten light cone realisations for each feedback model, rescaled to match the coverage of CFHTLenS (154 deg$^2$), and averaged over the five feedback models.
The band is centred  over the \textsc{AGN 8.5} model. The sample variance among the different AGN models varies by less than $\sim$20\%, so the grey band gives a good estimate of the standard error on the mean expected from CFHTLenS. The red dashed-dotted lines represent the 1-$\sigma$ confidence interval on the measured $\xi_{y\kappa}$ based on using different tSZ maps \cite{CFHT1}. The range includes statistical uncertainties as well systematic uncertainties due to foreground residuals in the tSZ maps (see \cite{CFHT1} for details).

To first order, the simulations match the amplitude and scale dependence of the measurements relatively well.  The two panels of Fig.~\ref{fig:Corr_y-kappa_smooth} show that measurement uncertainties contribute most to the overall error budget, followed by uncertainty in the cosmological parameters, while the uncertainty due to the different galactic feedback models is relatively small on these angular scales.  We find that with the current uncertainties, the model predictions from both \textit{Planck} and \textit{WMAP7} cosmologies are consistent with data.  Our results are also consistent with the findings of previous studies (e.g., \cite{McCarthy2014}, see also \cite{PlanckXXI}) that compared models with the {\it Planck} team's $C_{\ell}^{yy}$ measurement (the tSZ auto-correlation spectrum; \cite{PlanckXXI}).

As demonstrated  in Fig.~\ref{fig:ell_min_comparison_halo-model}, there is a very good agreement between the predictions from our fiducial AGN and halo model. This is due to the fact that, at high masses (which dominate the signal, see the decomposition by mass below), the fiducial AGN model has a pressure profile that is very similar to that derived from X-ray observations (see \cite{McCarthy2014}) and upon which the UPP was based.  However, one would expect deviations of the halo model predictions when contributions from baryons outside the virial radius, or, only low-mass groups and clusters are considered. A detailed study of such differences is left for future.

We now proceed to decompose the cross-correlation signal into the contributions from different redshifts, halo mass ranges and central distances (in units of the virial radius), analogous to that done previously for the tSZ auto-correlation by, e.g., \cite{Battaglia2012,McCarthy2014}.  A similar decomposition was performed in \cite{CFHT2} using the halo model, but the current hydrodynamical simulations incorporate many more astrophysical effects and are thus to be preferred.  In \cite{McCarthy2014}, it was shown that massive halos ($M_{200} \gtrsim 10^{14} \msun$) and small scales ($r  \lesssim R_{200}$) dominate the contributions to the $C_{\ell}^{yy}$ power spectrum.  Here we apply the same methodology to decompose the tSZ cross-correlation with weak lensing.

\begin{figure}[th]
\includegraphics[width=1.0\columnwidth]{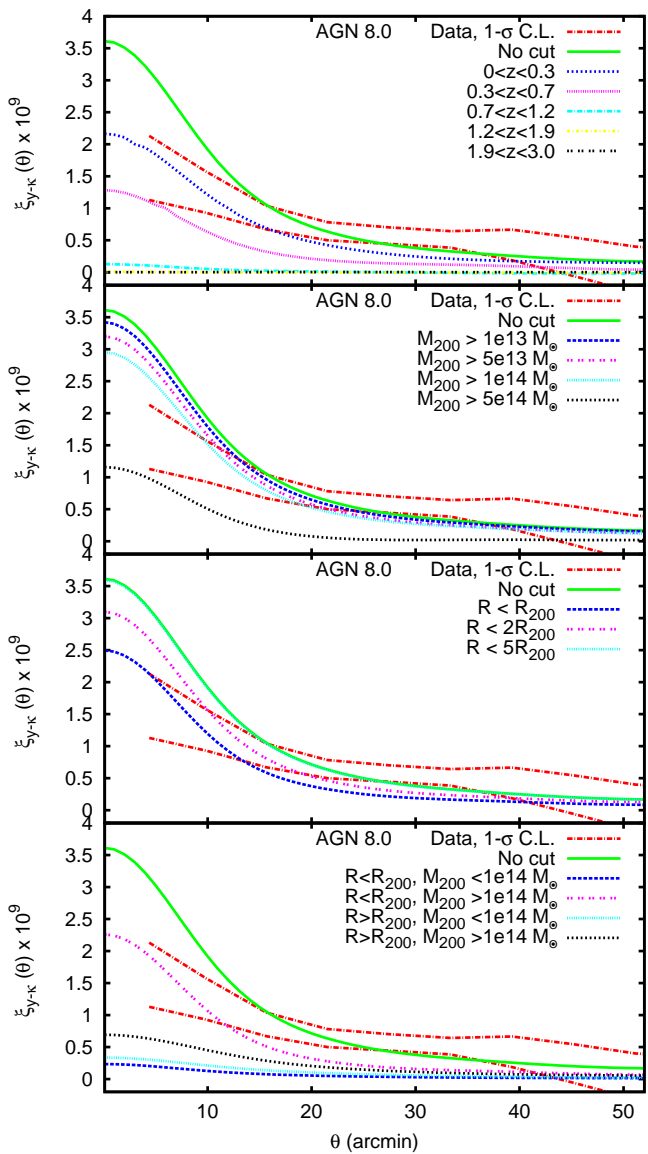}
\caption{Variations in the cross-correlation function, $\xi^{y\kappa}$, as a function of halo mass cut (top), central radial cut (middle), and both (bottom).  All panels assume the fiducial \textsc{AGN} feedback model and {\it Planck} cosmological parameters.  The 1-$\sigma$ confidence interval on the data \cite{CFHT1} is shown for a comparison.  Similar trends exist for the other feedback models (see text).  The majority of the signal is from massive halos, but low-mass halos still contribute $\sim$20\%. In contrast, most of the signal is from the hot gas within the virial radius of clusters, while relatively cold gas far from halo centers ($ r \gtrsim 5R_{200}$) contributes negligibly.}
\label{fig:Corr_y-kappa_smooth_Planck_cut_AGN}
\end{figure}

Fig.~\ref{fig:Corr_y-kappa_smooth_Planck_cut_AGN} shows the relative contributions to $\xi_{y\kappa}$ using different redshift cuts in the $y$ signal (top panel), halo mass cuts (second panel), different radius cuts (third panel), and selected combined cuts (bottom panel).  For simplicity, we only present cut results for the fiducial AGN baryonic feedback model ({\it Planck} cosmology), but the other feedback models show the same trends.  

Note that two additional cuts are implicitly imposed when examining the break down by halo mass and/or radius; namely that we only include haloes with a mass of at least $M_{200} \ge 10^{12}$ $\msun$ (we cannot examine lower mass haloes due to the finite resolution of the simulations) and we also impose a maximum radius cut of $5R_{200}$.  Some choice for the maximum radius is required when deciding which gas particles are associated with a particular halo and we wanted a radius that was sufficiently large to incorporate the virial shock region.  Below we show that these implicit cuts are inconsequential, as virtually all of the derived cross-correlation signal (summing all gas particles in the simulation volume, whether or not they are associated with haloes) can be accounted for by summing the contributions from gas associated with haloes with $M_{200} \ge 10^{12}$ $\msun$ and within $r < 5R_{200}$.

The top panel of Fig.~\ref{fig:Corr_y-kappa_smooth_Planck_cut_AGN} shows that $\xi_{y\kappa}$ is mostly from halos in $0 < z < 0.3$ and $0.3 < z < 0.7$ range\footnote{Note that the redshift intervals were chosen to divide the line of sight comoving distance back to $z=3$ into five equal segments.}.  This is what we expect as the redshift distribution of the lenses in CFHTLenS peaks around $z=0.4$ and $\xi_{y\kappa}$ would be sensitive to halos in that redshift range. The second panel of Fig.~\ref{fig:Corr_y-kappa_smooth_Planck_cut_AGN} shows that $\xi_{y\kappa}$ is dominated by halos with $M_{200} \gtrsim 10^{14}$ $\msun$, and that roughly half of the signal originates in halos with $M_{200} \gtrsim 5\times 10^{14}$ $\msun$.  It is interesting to note that halos with less than $10^{14}$ $\msun$ still contribute $\sim$20\% of the signal.  This supports previous findings in \cite{CFHT2} that low-mass halos produce a non-negligible fraction of the cross-correlation signal.  

The remaining two panels of Fig.~\ref{fig:Corr_y-kappa_smooth_Planck_cut_AGN} show the effects of radius cuts and of combined mass and radius cuts.  The middle panel demonstrates that most of the tSZ-lensing signal is from the hot gas within the virial radius of clusters and that contribution from the relatively cold gas far away from halo centers is ($ r \gtrsim 5R_{200}$) is negligible.  In the bottom panel, contributions are divided into four bins by mass {\em and} central radius: low mass ($10^{12} \msun \le  M_{\rm{halo}}  \le 10^{14} \msun$), high mass ($10^{14} \msun \le  M_{\rm{halo}}  \le 10^{16} \msun$), inner ($0 \le  r  \le R_{200} $) and outer ($R_{200} \le  r  \le 5R_{200} $). As expected, the high mass, inner bin produces the biggest fraction of the signal, but still only $\sim 50\%$ of the total.  It is clear that gas in the other regimes produces a considerable fraction of the signal in this fiducial \textsc{AGN} feedback model. 

Similar trends hold for the other feedback models, as the effects of feedback are generally small on the large scales probed here. (We discuss trends on smaller scales in the next section.)  The minor differences that are present are due to mechanisms that change the density and temperature of the gas near the center of clusters. For the \textsc{NOCOOL} model, there is more high temperature gas which results in a higher signal. In the \textsc{REF} model, feedback is inefficient so a large fraction of halo baryons are able to cool and form stars.  This reduces the gas fraction \cite{LeBrun2014} which lowers the tSZ amplitude.  An even lower signal is obtained when AGN feedback is introduced, with the suppression becoming greater as the AGN heating temperature is increased.  In this case, the reduction is not due to star formation, but rather ejection of gas from dark matter halos, but again, this is most pronounced on scales of a few arcmin, as discussed below.

A particular question of interest is whether or not the $\xi_{y\kappa}$ signal provides a useful probe of the ``missing baryons.''  To this end, it is important to note that the fractional contribution to the total $\xi_{y\kappa}$ signal in a given bin is not a measure of the fractional baryon density in that bin.  This is because the tSZ signal is proportional to the product of the gas density and its temperature and in general the gas is not isothermal.  In \cite{CFHT2} the baryon fraction was calculated analytically from the halo model using the simplifying assumption of isothermal gas. In this study, we extract gas information directly from the simulations, independent of its thermal state.  In particular, it is straightforward to sum the mass of hot gas using the same halo mass and central radius cuts used above.  This was done on the full 3D simulation at $z=0$ and the fractional contribution to $\Omega_{\rm b}$ was calculated simply by dividing the gas mass by the simulation volume and scaling to the critical density.  (We note that this decomposition depends only weakly on redshift over the range of redshifts we probe.)

Table \ref{table:contributions} compares the baryon fraction and the signal fraction in each of the combined bins based on the \textsc{AGN} feedback model with {\it Planck} cosmological parameters.  An interesting result emerges: the high mass, inner radius bin, which dominates the contribution to the signal, contains the lowest fractional contribution to the overall baryon density.  In contrast, the largest baryon fraction resides in the low mass, outer radius bin, which produces $\sim$11\% of the cross-correlation signal.  The high baryon fraction in this bin is due to the shape of the halo mass function -- there are many more low-mass halos than high-mass ones -- and to the fact that there is much more volume beyond $R_{200}$ than within it.

Note that the cumulative baryon density in Table~\ref{table:contributions} only accounts for 41\% of $\Omega_{\rm b}$.  This implies that the remaining $\sim$60\% of baryons reside in halos with masses less than $10^{12} \msun$ and/or at radii exceeding $5 R_{200}$ (i.e., part of the intergalactic medium).  These baryons make up most of the missing baryons, and this demonstrates that current tSZ-weak lensing measurements are not yet capable of detecting them. Future observations might be able to find them, for instance, by masking out all known halos and measuring the residual cross-correlation.

\begin{table}[th]
\centering
\caption{Fractional signal and baryon contributions by mass and radius for the \textsc{AGN} feedback model with {\it Planck} cosmological parameters. The mass bins are: low ($10^{12} \msun \lesssim  M_{\rm{halo}}  \lesssim 10^{14} \msun$), high ($10^{14} \msun \lesssim  M_{\rm{halo}}  \lesssim 10^{16} \msun$), and the radius bins are: inner ($0 \lesssim  r  \lesssim R_{200} $), and outer ($R_{200} \lesssim  r  \lesssim 5R_{200} $) radii.} 

\begin{tabular}{| l | c c|c|}
\hline
\multicolumn{1}{|l|} {} &  \multicolumn{2}{ |c| }{Simulation}  \\ 
\hline
Bin & Signal &~~~~ Baryon\\
\hline
Low mass, inner radii & 7$\%$ & 6$\%$\\
\hline
Low mass, outer radii& 11$\%$ & 24$\%$\\
\hline
High mass, inner radii & 55$\%$ & 4$\%$ \\
\hline
High mass, outer radii & 24$\%$ &7$\%$ \\
\hline
\end{tabular}
\label{table:contributions}
\end{table}

\subsection{Baryon feedback on small scales}
\label{smallscale}

Motivated by the results in \S\ref{largescale}, we explore the potential for higher-resolution $\xi_{y\kappa}$ measurements to discriminate among different feedback models.  Note that sub-arcminute lensing is already accessible with current lensing surveys (e.g. CFHTLenS as shown in \cite{JHD14}), and similar tSZ resolution will become more common with surveys like ACT and SPT.  In the following simulation analysis, neither the $\kappa$ nor $y$ maps are smoothed, allowing us to focus on the sub-arcminute cross-correlation signal.

\begin{figure*}[th]
\centering
\includegraphics[width=2.1\columnwidth]{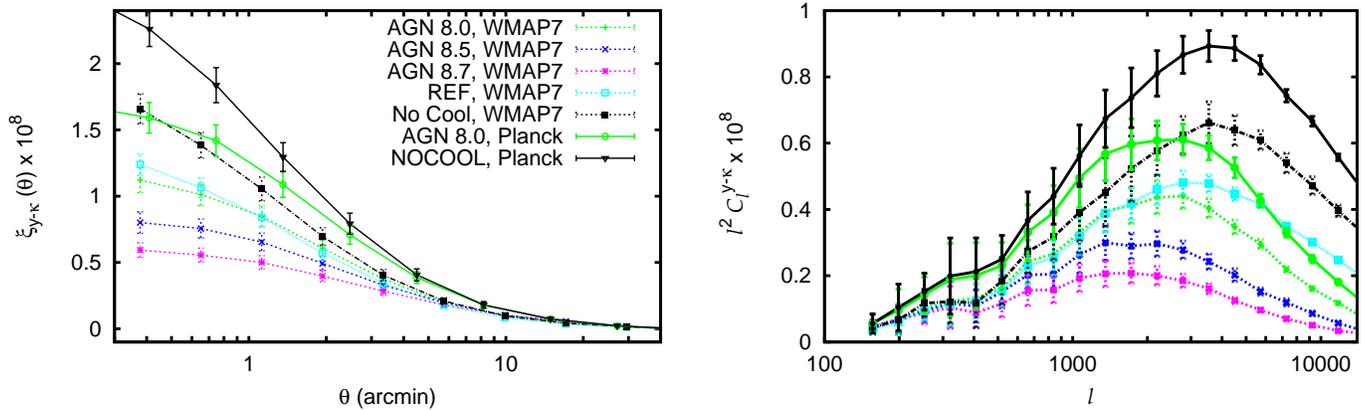}
\caption{Predicted $y$-$\kappa$ cross-correlation functions (left) and angular power spectra (right) from simulations with a range of feedback models.  Small angular scales are emphasized here, in contrast to Figure~\ref{fig:Corr_y-kappa_smooth}.  Predictions from two sets of cosmological parameters are shown, but only two feedback models are shown for the {\it Planck} parameters since the trends are the same in both cosmologies. The error bars represent the error on the mean from ten light cone realisations.  Baryon feedback is important on small scales and it dominates the uncertainty in the predicted cross-correlation.  Cosmological parameters also have a major impact; for example, the \textsc{AGN 8.0} model with {\it Planck} parameters is roughly equivalent to the \textsc{NOCOOL} model with {\it WMAP-7} parameters.}
\label{fig:Corr_y-kappa}
\end{figure*}

Fig~\ref{fig:Corr_y-kappa} shows several $y$-$\kappa$ cross-correlation functions (left) and angular power spectra (right) derived from the simulations for the feedback models considered above (for simplicity, only two feedback models are shown using {\it Planck} cosmological parameters since the feedback trends are similar in both cosmologies). The error bars represent the error on the mean value from ten light cone realisations.  As expected, the predicted signal is significantly higher using {\it Planck} parameters instead of {\it WMAP}-7 parameters.  This is due to the higher values of \{$\Omega_{\rm b}, \Omega_{\rm m}, \sigma_8$\} that {\it Planck} obtains, \{0.3175, 0.0490, 0.834\} vs. \{0.272, 0.0455, 0.81\}, which leads to the formation of more massive halos and hence a larger tSZ signal.  We note that these differences would be somewhat smaller with {\it WMAP}-9 parameters: \{0.288, 0.0472, 0.830\}.

Comparing the various feedback models, we see large differences in the predicted cross-correlation on scales of a few arc-minutes and smaller.  The \textsc{NOCOOL} model predicts the highest signal because there is no cooling or feedback mechanism in this model; consequently, the hot gas roughly tracks the dark matter.  This leads to a relatively high density of hot gas at the center of halos.  In the \textsc{REF} model, cooling, star formation and SN feedback (which is generally inefficient at these mass scales) are included.  This lowers the predicted tSZ amplitude, as a large fraction of baryons are converted into stars.  When AGN feedback is added, low-entropy gas is ejected from the halo (as opposed to forming stars) lowering the tSZ amplitude even further.  This process becomes increasingly important as the feedback heating temperature is increased. 

\begin{figure}
\centerline{\includegraphics[width=1.05\columnwidth]{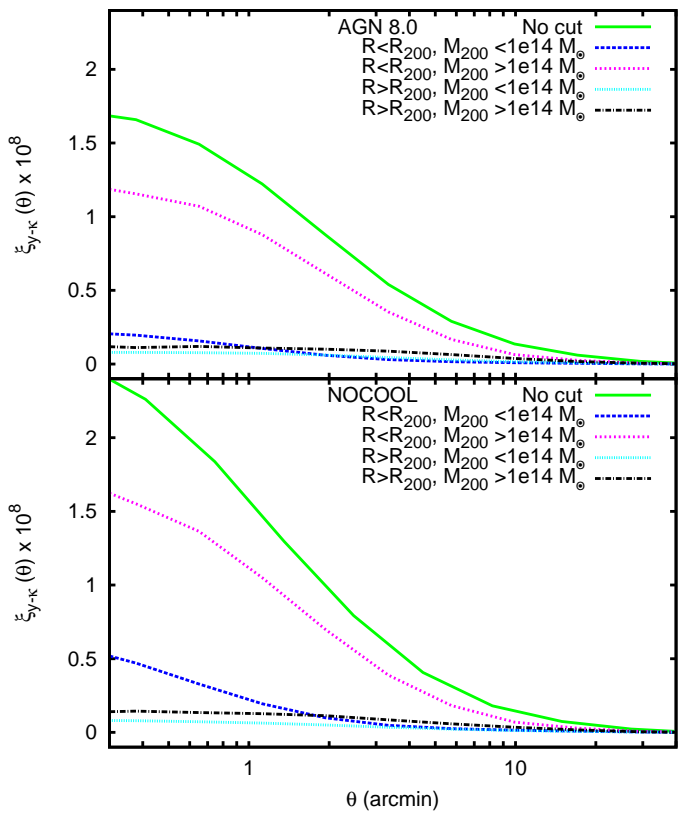}}
\caption{Decomposition of $\xi_{y\kappa}$ by halo mass and central radius bins for the {\textsc{AGN}} (top) and \textsc{NOCOOL} (bottom) feedback models, using {\it Planck} cosmological parameters.  At the center of clusters (the smallest scales probed here), baryon feedback determines not only the amplitude of the tSZ-lensing signal, but also the relative contributions from halos with different masses. For example, the \textsc{NOCOOL} model retains hot gas in low-mass halos, thus boosting the relative strength of the low mass, inner radius bin in that model.}
\label{fig:Corr_y-kappa_Planck_rmcut_p2}
\end{figure}

Fig.~\ref{fig:Corr_y-kappa_Planck_rmcut_p2} shows the decomposition of $\xi_{y\kappa}$ by mass and radius on small scales for the \textsc{NOCOOL} and \textsc{AGN} feedback models.  As before, the majority of the signal originates from within the virial radius of massive halos.  However, the next leading term on these scales is from within the virial radius of low-mass halos.  This does not contradict the findings of Section \ref{largescale} on large scales, as the gas at the center of low-mass halos is hot and would produce more tSZ signal than the cooler gas in the outskirts of massive halos.   These low-mass halos are simply not resolved on the scale of {\it Planck}'s angular resolution.

Table~\ref{table:contributions_small} gives the fractional contributions to $\xi_{y\kappa}$ from the inner radius bins ($r \lesssim R_{200}$) for both the \textsc{AGN} and \textsc{NOCOOL} models (note again that no smoothing has been applied to the simulated maps here).  It also tabulates the fractional contributions to $\Omega_{\rm b}$ from each bin, as in Table~\ref{table:contributions}.  Again, a relatively small fraction of baryons ($\sim$5\%) produce most of the cross-correlation signal. A combination of high resolution tSZ maps with large-area weak lensing surveys could provide a wealth of information about the state of gas in groups and clusters. 

\begin{table}[th]
\centering
\caption{The fractional signal and baryon contributions from within $R_{200}$ for the \textsc{AGN} and \textsc{NOCOOL} models using {\it Planck} cosmological parameters. The mass bins are the same as in Table \ref{table:contributions}, but the simulated maps are not smoothed here.} 
\begin{tabular}{| l | c c|c c|}
\hline
\multicolumn{1}{|l|}{} &  \multicolumn{2}{ |c| }{\textsc{AGN}} &  \multicolumn{2}{ |c| }{\textsc{NOCOOL}} \\ 
\hline
Cut & Signal &~~~ Baryon& Signal &~~~ Baryon \\ 
\hline
Low mass, inner radii & 11$\%$     & 6$\%$  & 18$\%$ & 14$\%$\\
\hline
High mass, inner radii & 73$\%$  &  4$\%$ & 72$\%$  & 6$\%$\\
\hline
\end{tabular}
\label{table:contributions_small}
\end{table}

\section{Summary}
\label{sec:summary}

We have analyzed tSZ and gravitational lensing maps derived from the cosmo-OWLS hydrodynamical simulations to interpret the measured $\xi_{y\kappa}$ correlation function reported in \cite{CFHT1}.  We find relatively good agreement between the predicted and measured signals on the angular scales probed by the data.  Note that the predictions depend on the choice of background cosmology but, given the present statistical and systematic uncertainties (in both the observations and simulations), both the {\it WMAP} and {\it Planck} cosmologies provide reasonable fits.

The cosmo-OWLS simulations confirm a previous finding \cite{CFHT2} that $\sim$20\% of the cross-correlation signal arises from low-mass halos, and about a third of the signal originates in diffuse gas beyond the virial radius, up to $r \sim 5R_{200}$.  A majority of the signal comes from a small fraction of baryons within halos ($r \lesssim R_{200}$), while about half of all baryons reside outside ($r \gtrsim 5R_{200}$) and are too cool ($T \sim 10^5 K$) and rarefied to contribute significantly to the cross-correlation signal. In detail there are small differences in the predicted breakdown by radius and mass between the halo model and simulations, which are plausibly due to the neglect of feedback on the total mass profile and the pressure distribution of the hot gas, particularly for galaxy groups where the effects of AGN feedback are substantial.

Two factors limited the study of cosmology with tSZ-lensing correlations in \cite{CFHT1}.  First, the relatively low angular resolution of the {\it Planck} tSZ maps precluded exploring arc-minute scales.  Baryon feedback is important on small scales and it can significantly affect the tSZ amplitude.  Higher resolution tSZ maps from ACT, SPT and others will enable us to distinguish feedback models and constrain baryonic processes inside and around halos.  Second, the relatively small ariel coverage of CFHTLenS restricts our ability to firmly distinguish between {\it WMAP}-7 and {\it Planck} cosmological parameters. The errors on $\xi_{y\kappa}$ will shrink significantly with several upcoming weak lensing surveys such as RCSLenS, KiDS and DES, which cover $\sim$50 times more sky area than CFHTLenS. This will help distinguish cosmological models on large scales and baryon feedback models on small scales.  Further, better CMB data from the final release of {\it Planck} can help characterize systematic uncertainties in the tSZ maps. 

An interesting direction for future work would be to examine feedback processes in different galaxies.  For example, measuring $y$-$\kappa$ correlations in classes of objects, rather than cross-correlating maps, could provide useful information about gastrophysics e.g. in clusters and groups of galaxies as a function of mass and redshift.  This is currently work in progress.

\section{acknowledgments}
We would like to thank the members of the OWLS team for their contributions to the development of the simulation code used here.  We thank Gilbert Holder, Joop Schaye, Jessica Ford, Tilman Troester, and Hideki Tanimura for useful discussions.  AH is supported by NSERC postdoctoral fellowship. IGM is supported by a STFC Advanced Fellowship at Liverpool JMU. JHD, LVW, and GH are supported by NSERC and the Canadian Institute for Advanced Research.

This paper made use of the Planck Legacy Archive (\url{http://archives.esac.esa.int/pla}) and the Canada-France-Hawaii Telescope Lensing Survey (\url{http://www.cfhtlens.org/}).

Before submitting this manuscript, we became aware of similar
calculations presented in~\cite{Battaglia2014}, and we subsequently
exchanged correspondence with these authors. We leave a detailed
comparison of the results for an updated analysis.

\appendix

\section{Light cone extraction and limitations of viewing angle}
\label{lowell}

The light cones produced from simulations provide a viewing angle of 
\begin{equation}
\theta_{max} = \frac{L_{box}}{D(z_{max})}
\end{equation}
where $L_{box}$ is the simulation box size (in $Mpc/h$) and $D$ is the comoving line-of-sight distance to the maximum redshift of the simulation ($z=3$). In our case $L_{bpx} = 400$, yielding $\theta_{max} \approx 5^{\circ}$ for a $\Lambda$CDM model with current cosmological parameters.  This corresponds to a minimum Fourier-space multipole of $\ell_{min} = 72$.  The simulation-based maps are, therefore, missing modes below $\ell_{min}$ which could have an effect on real space cross-correlation function (i.e., underestimate the power at a given angle, $\theta$).

\begin{figure}[th]
{\centering{
\includegraphics[width=1.0\columnwidth]{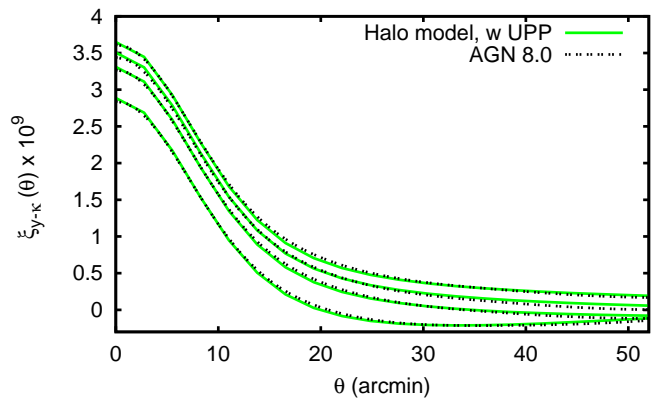}
}}
\caption{Prediction of the halo model (with UPP) (Solid lines), and results from \textsc{AGN 8.0} simulations (dashed lines) for $\xi_{y\kappa}$ when different $\ell_{min}$ cuts are applied. From top to bottom, the $\ell_{min}$ cuts are $1, 50, 100$ and $200$. Missing low-$\ell$ modes leads to a loss of power on all scales. There is a good agreement between $\xi_{y\kappa}$ predictions from halo model and those computed from simulated maps.  }
\label{fig:ell_min_comparison_halo-model}
\end{figure}

In Fig.~\ref{fig:ell_min_comparison_halo-model} we show the halo model (with UPP; green curves) prediction for $\xi_{y\kappa}$ where different $\ell_{min}$ are imposed. One notes that cutting low-$\ell$ leads to a loss of power on all scales in real space.  We correct for the effect of the missing modes on the simulation-based maps by assuming that the $\ell~C_{\ell}^{y\kappa}$ approaches a plateau (constant) below $\ell=100$.  This choice is motivated by the results of the halo model as well as experiments with larger viewing angles (but smaller $z_{max}$ values).  Thus, our real space cross-correlation functions from the simulations are computed by using the Fourier space cross-spectra for $\ell \ge 100$ directly from the simulations and imposing a constant below $\ell = 100$.  The conversion to real space is done via Legendre transformation, as described in \cite{CFHT2}.

The dashed black curves in Fig.~\ref{fig:ell_min_comparison_halo-model} show the resulting real space cross-correlation functions for the \textsc{AGN} model.  There is remarkably good agreement with the predictions of the halo model for a given minimum multipole, simultaneously demonstrating the accuracy of our correction procedure as well as the agreement of the halo model with the fiducial AGN simulation.  Note that for the cases with $\ell_{min} = 100,200$ no extrapolation of the simulation cross-spectra is necessary before transforming to real space.

\end{document}